\documentclass[11pt,a4paper]{article}

\usepackage{amsmath}
\usepackage[T1]{fontenc}
\usepackage[utf8]{inputenc}
\usepackage{amsfonts}
\usepackage{amsthm}
\usepackage{amstext}
\usepackage{amssymb}
\usepackage{color}
\usepackage{bbold}
\usepackage{hyperref}
\usepackage{url}

\newcommand{\rs}{\alpha}
\newtheorem{remark}{Remark}

\begin{document}

\title{Infall time 
in the Eddington-Finkelstein metric, with application to Einstein-Rosen bridges}

\author{Pascal Koiran\\
Ecole Normale Sup\'erieure de Lyon, France\thanks{Email: pascal.koiran@ens-lyon.fr}}

\maketitle

\begin{abstract}
 The  Eddington-Finkelstein metric is obtained from the  Schwarzschild metric by a change of the
time variable. It is well known that a test mass falling into a black hole does not reach the event horizon for any finite value of the Schwarzschild time variable~$t$. By contrast, we show that 
the event horizon is reached  for a finite value of the Eddington-Finkelstein time variable $t'$.
Then we study in Eddington-Finkelstein time the fate of a massive  particle traversing an Einstein-Rosen bridge and obtain a different conclusion than recent proposals in the literature: 
we  show that the particle reaches the wormhole throat for a finite value~$t'_1$
of the time marker~$t'$, and continues its trajectory across the throat for $t'>t'_1$.
Such a behavior does not make sense in Schwarzschild time  since it would amount to continuing the trajectory of the particle "beyond  the end of time."
\end{abstract}

\section{Introduction}

The simplest black hole model is described by the  Schwarzschild metric:
\begin{equation} \label{schwarz}
ds^2=(1-\frac{\rs}{r})c^2dt^2-\frac{dr^2}{1-\rs/r}-r^2(d\theta^2+\sin^2 \theta d\phi^2),
\end{equation}
where $\rs$ is the Schwarzschild radius.
The singularity at the event horizon ($r=\rs$)  is an artificial one, due to a bad choice of coordinates. 
A better choice is provided by the Eddington-Finkelstein coordinates, where the time marker $t$ is replaced by:
\begin{equation} \label{EFtime}
t'=t+\frac{\alpha}{c} \ln \left| \frac{r}{\rs} -1 \right|.
\end{equation}
In the new coordinates $(t',r,\theta,\phi)$, the  metric takes the form 
\begin{equation} \label{EFmetric}
ds^2=(1-\frac{\rs}{r})c^2dt'^2-\frac{r+\alpha}{r}{dr^2}-\frac{2\rs c}{r}dr\, dt' - r^2(d\theta^2+\sin^2 \theta d\phi^2),
\end{equation}
which is indeed nonsingular for $r=\alpha$.\footnote{Throughout the paper we use the $(+, -, -, -)$ sign convention for the signature of the metric. This was the choice of most of the pioneers of relativity theory, including Einstein himself (as in e.g.~\cite{ER35}),  Schwarzschild~\cite{Schwarzschild16,Schwarzschild16b}, Weyl~\cite{weyl17}, Lemaître~\cite{lemaitre33}, Eddington~\cite{Eddington24} or Finkelstein~\cite{finkelstein58}.}

%changement par rapport à physics letters B:
  It is well known that an infalling mass following a radial trajectory 
in the Schwarzschild spacetime reaches the Schwarzschild radius $r=\alpha$ in finite proper time;
but the  Schwarzschild radius is not reached for any finite value of the time parameter $t$. In this paper we show that the opposite is true in the Eddington-Finkelstein metric~(\ref{EFmetric}): the infalling mass will reach 
the Schwarzschild radius for a finite value of the time parameter $t'$. This effect is not directly observable 
because a light ray emitted by the particle at the Schwarzschild
radius will not reach a distant observer in finite time.

Nevertheless, the finiteness of infall time will
play an important role in Section~\ref{bridges}, where we analyze the fate of a massive particle
traversing an Einstein-Rosen bridge. This question was considered in several recent papers~\cite{poplawski10,katanaev14,guendelman17}. In particular, for an observer located across
from the wormhole throat,  it was suggested that:
\begin{itemize}
\item[-] the throat appears as a past black hole horizon~\cite{guendelman17}.
\item[-]  the particle goes "backward in time" after crossing the throat, and the observer sees an antiparticle falling into 
a %Schwarzschild
 black hole~\cite{katanaev14}.
\end{itemize}
In this paper we study a version of the Einstein-Rosen bridge obtained by coupling the  ingoing Eddington-Finkelstein metric~(\ref{EFmetric}) to the outgoing version of this metric.
Our analysis leads to a different conclusion than in~\cite{guendelman17,katanaev14}: the particle reaches the throat for a finite value $t'_1$
of the time marker $t'$, and continues its trajectory across the throat for $t'>t'_1$. Such a behavior
clearly does not make sense if one works in Schwarzschild time, since it would amount to 
continue the particle's trajectory "beyond the end of time."

\section{Trajectories of light rays} 
 \label{geo}

In Section~\ref{sec:mass} we study the trajectories  of massive particles in the Eddington-Finkelstein metric. 
As a preliminary to this study, we consider in this section the trajectories of  light rays.
In contrast to the Schwarzschild metric, we will see that in the Eddington-Finkelstein metric the  speed of light $dr/dt'$ is different for inward and outward pointing light rays. 
This effect is of course not physically observable: it is only an artifact of the choice of a system of coordinates.
 Measuring the one-way speed of light involves indeed the use of synchronized clocks~\cite{will92}. The proper times measured by these
 clocks will be the same in both metrics since they are equivalent up to a change of variables. 

We only consider radial trajectories, so~(\ref{EFmetric}) simplifies to:
\begin{equation} \label{EFrad}
ds^2=(1-\frac{\rs}{r})c^2dt'^2-\frac{r+\alpha}{r}{dr^2}-\frac{2\rs c}{r}dr\ dt'.
\end{equation}

Setting $ds^2=0$ in~(\ref{EFrad}), we obtain for $dr/dt'$ the quadratic equation:
$$(r+\alpha) (dr/dt')^2+{2\rs c} (dr/dt')-(r-{\rs})c^2 =0$$
which admits the two solutions $-c$ for inward rays and $\frac{r-\rs}{r+\rs} c$ for outward rays. 
We therefore have a smaller speed of light for outward rays. 

Since the speed of inward light rays is constant and equal to $c$, they will  reach the Schwarzschild radius $r=\rs$ for a finite value of the time marker~$t'$. This contrasts with the Schwarzschild  metric~(\ref{schwarz}), for which the amount of time $\Delta t$ needed to reach the Schwarzschild radius 
is infinite.
For outward rays we have $cdt'/dr=(r+\rs)/(r-\rs)$. The time needed to travel from 
radial position $r=R$ to $r=R_0>R$ is equal to:
$$\Delta t' = \int_R^{R_0} \frac{r+\rs}{r-\rs} \frac{dr}{c}.$$
We therefore have a logarithmic divergence as $R \rightarrow \alpha$: escape time is infinite like in the 
Schwarzschild  metric. We will derive similar results for massive particles (finite infall time but infinite escape time)
in the next section.
Toward this study we note that 
\begin{equation} \label{outineq}
c\frac{dt'}{dr} > \frac{r+\rs}{r-\rs}
\end{equation}
for outward moving massive particles since $c{dt'}/{dr} = \frac{r+\rs}{r-\rs}$ for outward light rays.

\section{Trajectories of massive particles} 
 \label{sec:mass}

In this section we study radial trajectories  of massive particles in the Eddington-Finkelstein metric. 
In order to compute geodesics in metric~(\ref{EFrad}) we consider the 
 associated Lagrangian
$$F=\sqrt{(1-\frac{\rs}{r})c^2\dot{t'}^2-\frac{r+\alpha}{r}{\dot{r}^2}-\frac{2\rs c}{r} \dot{r} \dot{t'}}$$
where $\dot{r}=dr/ds$, $\dot{t'}=dt'/ds$.
By the Euler-Lagrange equations, geodesics must satisfy the differential equation:
$$\frac{d}{ds}\left( \frac{\partial F}{\partial \dot{t'}} \right) = \frac{\partial F}{\partial {t'}}.$$
Since $F$ is independent of $t'$ and $F$ is identically equal to 1 along the geodesic we obtain
\begin{equation} \label{mueq}
(1-\frac{\rs}{r})c\dot{t'}-\frac{\rs}{r}\dot{r}=\mu
\end{equation}
where $\mu$ is a constant of motion. Or equivalently, 
\begin{equation} \label{tpoint}
c\dot{t'} = \frac{\mu r + \alpha \dot{r}}{r-\alpha}.
\end{equation}
Note that $\mu>0$ for an infalling trajectory since (\ref{mueq}) expresses $\mu$ as the sum of two positive terms.
We will see in a moment that outward moving particles also satisfy the same constraint 
$\mu>0$.

We need a second equation in order to compute geodesics. The constraint $F=1$ gives:
\begin{equation} \label{quadreq}
(1-\frac{\rs}{r})c^2\dot{t'}^2-\frac{r+\alpha}{r}{\dot{r}^2}-\frac{2\rs c}{r} \dot{r} \dot{t'}=1.
\end{equation}
From this quadratic equation in $\dot{r}$ we could express $\dot{r}$ as a function of $r$ and~$\dot{t'}$.
We will instead express $\dot{r}$ as a function of $r$ only.
For this we replace $c\dot{t'}$ in~(\ref{quadreq}) by the right-hand side of (\ref{tpoint}).
This yields another quadratic equation:
$$\dot{r}^2-\mu^2+1-\rs/r=0$$
which admits the solution 
\begin{equation} \label{eq:rdot}
 \dot{r}=\nu \sqrt{\mu^2-1+\rs/r}
 \end{equation}
where $\nu =-1$ for an infalling trajectory, $\nu =+1$ for an outgoing trajectory.
Hence $\mu \neq 0$ if there is a point on the trajectory with $r > \rs$.
By substitution of this solution in~(\ref{tpoint}) we can also express $\dot{t'}$ as a function of $r$.
Then we have $$dt'/dr=\dot{t'}/\dot{r} = (\nu/c)g(r)$$
where 
\begin{equation} \label{geq}
g(r)=\frac{\mu r + \rs \nu \sqrt{\mu^2-1+\rs/r}}{(r-\rs) \sqrt{\mu^2-1+\rs/r}}.
\end{equation}
Consider a portion of geodesic beginning at $t'=t'_0$ and $r=R_0 > \alpha$.
The radial position $r=R$ is reached at time $\displaystyle t'(R)=t'_0+\int_{R_0}^R (\nu /c) g(r) dr.$
In order to show that infall time to the Schwarzschild radius is finite, we will show that $g(r)$ remains bounded as $r \rightarrow \rs$ when $\nu = -1$.

The numerator of~(\ref{geq}) is indeed equal to $(r-\rs)\mu+\rs (\mu-\sqrt{\mu^2-1+\rs/r})$.
The quotient of the first term $(r-\rs)\mu$ by the denominator of~(\ref{geq}) converges to 1 as $r \rightarrow \rs$. The quotient of the second term by the same denominator is equal to:
$$\frac{\alpha}{r(\mu+\sqrt{\mu^2-1+\rs/r})}.$$
It converges to $1/(2\mu)$ as $r \rightarrow \alpha$ 
(recall indeed that $\mu>0$ for an infalling trajectory). We conclude that 
$\lim_{r \rightarrow \rs} g(r) = 1+1/{(2\mu)}$
is finite, as needed. 

Let us now  turn our attention to outgoing trajectories. 
We have with~(\ref{geq}) an exact expression for $dt'/dr$, but we also know 
that $dt'/dr$ must satisfy~(\ref{outineq}). From these two constraints one can easily
derive the inequality
$$\mu  > \sqrt{\mu^2-1+\rs/r},$$
which implies $\mu > 0$.
This implies in turn that for outgoing trajectories ($\nu=1$), $g(r)$ is equivalent to $2\rs/(r-\alpha)$ as $r \rightarrow \rs$.
As a result, $t'(R)$ exhibits a logarithmic divergence as $R \rightarrow \rs$: infall time is finite but 
escape time is infinite. 
\begin{remark} \label{muremark}
By~(\ref{eq:rdot}),  if $\mu < 1$ the particle's trajectory is limited to the region $r \leq \rs/(1-\mu^2)$. For $\mu \geq 1$ 
an outgoing trajectory will escape to infinity, and likewise for an ingoing trajectory: it can be continued backward in time
as the trajectory of a particle "plunging from infinity" into the black hole.
\end{remark}

\subsection{Finite infall time: an alternative derivation} \label{alter}

We now present an alternative derivation of the main result of Section~\ref{sec:mass}:
infall time for massive particles is finite.
For this we leverage the very standard
computation of geodesics in Schwarzschild spacetime~(e.g., \cite[chapter 6]{ABS} 
or \cite[chapter 25]{MTW}).
 Recall that for a radial geodesic in metric~(\ref{schwarz}) we have
\begin{equation} \label{schwarzradial}
(1-\alpha/r) \dot{t} = \ell,\ 1-\alpha/r = c^2\ell^2 - \dot{r}^2,
\end{equation}
where $\dot{r} = dr/ds$, $\dot{t}=dt/ds$ and $\ell$ is a constant of motion. 
From these equations we can compute $dt/dr$ as a function of $r$, and  $t$ as a function of $r$ by integration.
We can then express $t'$ as a function of $r$ from~(\ref{EFtime}),
and it remains to show that $t'(r)$ remains bounded all the way to $r=\alpha$. 
Let us now fill in the details.
From~(\ref{schwarzradial}) we have
\begin{equation} \label{dtdr}
\frac{dt}{dr} = \frac{- \ell}{1-\alpha/r} \sqrt{\frac{r}{\alpha+r(c^2\ell^2-1)}}.
\end{equation}
Consider a portion of geodesic beginning at $t=t_0$ and $r=R_0 > \alpha$.
From~(\ref{EFtime}) we have 
$$ct'(R)=ct(R)+{\alpha} \ln(R/\alpha-1) = ct(R)+\alpha \ln(R_0/\alpha-1)-\int_{R}^{R_0} \frac{dr}{r/\alpha-1}.$$
Then  from~(\ref{dtdr}) we have
$\displaystyle ct'(R)=ct'_0+\int_R^{R_0} f(r) dr$
where
$$f(r)=\frac{1}{r-\alpha}\left[\frac{c\ell r^{3/2}}{\sqrt{\alpha+r(c^2\ell^2-1)}}-\alpha\right]$$
and $t'_0=t_0+(\alpha/c) \ln(R_0/\alpha-1)$.
This is the same expression that was given for $t'(R)$ earlier in Section~\ref{sec:mass} in terms of the integral of the function $g(r)$.
Indeed we have $f(r)=g(r)$ when the constants of motion satisfy  $\mu=c \ell$, and $\nu=-1$ in~(\ref{geq}).
In particular, we obtain the same value of the infall time as before. 

\subsection{The outgoing Eddington-Finkelstein metric}
\label{sec:out}

Besides the Schwarzschild metric, we have so far considered only the {\em ingoing} Eddington-Finkelstein metric~(\ref{EFmetric}).
There is also an  {\em outgoing} form of the latter metric~\cite[chapter 31]{MTW}, obtained from the Schwarzschild metric
by the change of variables 
\begin{equation} \label{outime}
t'=t-\frac{\alpha}{c} \ln \left| \frac{r}{\rs} -1 \right|.
\end{equation}
Both metrics are of the form
\begin{equation} \label{EFmetrics}
ds^2=(1-\frac{\rs}{r})c^2dt'^2-\frac{r+\alpha}{r}{dr^2}+\delta\frac{2\rs c}{r}dr\ dt' - r^2(d\theta^2+\sin^2 \theta d\phi^2),
\end{equation}
where $\delta=-1$ for the ingoing metric and $\delta=+1$ for the outgoing metric.
These two metrics are images of each other by the time reversal $t \mapsto -t$. Ingoing trajectories in one metric
therefore correspond to outgoing trajectories in the other metric. As a result, for the outgoing metric escape time is finite
but infall time is infinite.\footnote{The same line of reasoning can be applied to the Schwarzschild metric.
Since it is invariant under time reserval, the fact that infall time is infinite follows from the knowledge that escape time is infinite, and vice versa.}
This conclusion can also be reached by a direct analysis of geodesics, or the  "alternative derivation" 
of Section~\ref{alter}.
In Section~\ref{bridges} we will pair these two metrics through an Einstein-Rosen bridge.

\section{Einstein-Rosen bridges in Eddington-Finkelstein time} \label{bridges}

In recent years, several papers~\cite{poplawski10,katanaev14,guendelman17} have studied the fate of particles traversing an Einstein-Rosen bridge.\footnote{The standard "textbook treatment"~\cite{FW62,MTW} of Einstein-Rosen bridges leads to the conclusion that they are non-traversable, but this analysis contains a serious flaw. This point is  clearly explained in the appendix of~\cite{guendelman10}, see also~\cite{guendelman17,poplawski10}.}
The Einstein-Rosen bridge was constructed in the seminal paper~\cite{ER35} from the Schwarzschild solution~(\ref{schwarz}) by the change of variables $r=\alpha+u^2$, where the new radial coordinate $u$ is allowed to take any real value. The resulting spacetime contains two copies of the exterior Schwarzschild region $r > \alpha$ glued together at the wormhole throat $r=\alpha$.

An alternative change of variables $r=\alpha+|\eta|$ was suggested in~\cite{guendelman10,guendelman17}.
Contrary to the original construction in~\cite{ER35},  the resulting spacetime satisfies the Einstein field equations everywhere including at the throat $\eta=0$ because a lightlike brane is added at the throat (see ~\cite{guendelman10,guendelman17} for details).
Here we propose to combine their change of variables with the Eddington-Finkelstein change of variables~(\ref{EFtime}).

We therefore have \begin{equation} \label{t'bridge}
t'=t+\frac{\alpha}{c} \ln \ \frac{|\eta|}{\rs},
\end{equation}
and $$dt'=dt+\alpha d \eta/c = dt - (\delta \alpha)/(c |\eta|)$$ where $\delta=-1$ in the region $\eta >0$, $\delta=+1$ in the region $\eta < 0$.
This amounts to applying the change of variables of the {\em ingoing}  Eddington-Finkelstein metric in the region $\eta >0$,
and the change of variables~(\ref{outime}) of the {\em outgoing} metric in the region  $\eta < 0$.
After substitution in~(\ref{schwarz}) we therefore obtain a spacetime described by the ingoing Eddington-Finkelstein metric
in the region $\eta > 0$, and by the outgoing  metric in the region $\eta < 0$.
The resulting line element is:
\begin{equation} \label{bridgeline}
ds^2=\frac{|\eta|}{\alpha+|\eta|}c^2dt'^2-\frac{2\rs+|\eta|}{\alpha+|\eta|}{d\eta^2}-\frac{2\rs c}{\rs+|\eta|}d\eta\ dt'.
\end{equation}
Here we have omitted the rotational degrees of freedom $\theta,\phi$ like in~(\ref{EFrad}).
This line element already appears in the appendix of~\cite{guendelman10} in slightly different notations.
The description in~\cite{guendelman17} of the fate of a particle traversing an Einstein-Rosen bridge
is based on~\cite{guendelman10}, but it is nevertheless significantly different from the description
that we now propose. The reason for this difference is that~\cite{guendelman17} works 
in Schwarzschild time instead of Eddington-Finkelstein time.

Let us consider a massive particle falling into the Einstein-Rosen bridge along a radial trajectory, beginning at 
time $t'=t'_0=0$ for a value of the radial parameter $\eta=\eta_0 > 0$ and a value $\tau=\tau_0=0$ of the particle's proper time.
It will reach the wormhole throat $\eta=0$ for a finite value $\tau_1$ of its proper time, 
and also (by the analysis in Section~\ref{sec:mass}) for a finite value $t'_1$ of the time marker $t'$.
For $\tau > \tau_1$ the parameter $\eta$ continues to decrease and the outgoing metric takes over.
The particle continues  in the region $\eta < 0$, where (by Section~\ref{sec:out}) its trajectory will be a mirror image of the trajectory in the region $\eta>0$. 
In particular, it will reach the position $\eta=-\eta_0$ for a value of its proper time $\tau=2\tau_1$ and at $t'=2t'_1$.
The particle's subsequent fate depends on the value of the constant of motion $\mu$. 
By Remark~\ref{muremark}, the particle will escape to infinity ($\eta \rightarrow -\infty$) 
if $\mu \geq 1$. For $\mu < 1$, the particle will reach a maximum value of $|\eta|$ and it will then 
fall back toward the wormhole throat %$\eta =0$ 
in finite proper time. 
Since we are now in the region where the outgoing metric applies, the throat $\eta=0$ will not be reached for any finite value of the time marker $t'$.

The above behavior is significantly different than in~\cite{guendelman17,katanaev14}: as recalled 
in the introduction, these authors have suggested that an observer located across the wormhole
throat sees it as a past black hole horizon~\cite{guendelman17}, or sees an antiparticle
falling into a 
 black hole~\cite{katanaev14}.

\section{Conclusions}

We have studied the fate of a massive particle traversing an Einstein-Rosen bridge, and have reached a different conclusion than previous proposals in the literature~\cite{katanaev14,guendelman17}:  the particle reaches the wormhole throat for a finite value~$t'_1$
of the time marker $t'$, and continues its trajectory across the throat for $t'>t'_1$. This difference is due to our use of Eddington-Finkelstein time as our time marker $t'$. The Eddington-Finkelstein metric was originally conceived as a tool for the study of the 
Schwarzschild metric. The results presented in this paper suggest that it could also be considered as an {\em alternative} 
to the Schwarzschild metric for the study of Einstein-Rosen bridges. Indeed, working in Schwarzschild time leads to
the different conclusions that an observer located across the wormhole
throat sees it as a past black hole horizon~\cite{guendelman17}, or sees an antiparticle
falling into a 
 black hole~\cite{katanaev14}.
 
It is a central tenet of general relativity that metrics which are equivalent 
 up to a change of variables should be considered physically equivalent.  This is not in contradiction with our proposal because the change of variables~(\ref{t'bridge}) leading from Schwarzschild to Eddington-Finkelstein time is well-defined
 only when the radial parameter $\eta$ of the Einstein-Rosen bridge takes a nonzero value;
whereas in the line element~(\ref{bridgeline}) any real values of $t'$ and $\eta$ are allowed.

\end{document}